\documentclass[twocolumn,
prl
,superscriptaddress
]{revtex4}

\usepackage{subfloat,float}
\usepackage[caption=false]{subfig}
\usepackage{amssymb}
\usepackage{amsmath}
\usepackage[pdftex]{graphicx}
\usepackage{epstopdf}
\usepackage{color}

\begin{document}

\author{Marianne Bauer}
\email{Marianne.Bauer@physik.uni-muenchen.de}
\affiliation{%
Arnold Sommerfeld Center for Theoretical Physics and Center for NanoScience, Department of Physics, Ludwig-Maximilians-Universit\"at M\"unchen, Theresienstr.~37, D--80333 Munich, Germany}
\author{Erwin Frey}
\email{frey@lmu.de}
\affiliation{%
Arnold Sommerfeld Center for Theoretical Physics and Center for NanoScience, Department of Physics, Ludwig-Maximilians-Universit\"at M\"unchen, Theresienstr.~37, D--80333 Munich, Germany}

\title{Delays in fitness adjustment can lead to coexistence \\ of hierarchically interacting species}

\begin{abstract}
	Organisms that exploit different environments may experience a stochastic delay in adjusting their fitness when they switch habitats. 
	We study two such organisms whose fitness is determined by the species composition of the local environment, as they interact through a public good. 
	We show that a delay in fitness adjustment can lead to coexistence of the two species in a metapopulation, although the faster growing species always wins in well-mixed competition experiments. 
	Coexistence is favored over wide parameter ranges, and is independent of spatial clustering.
	It arises when species are heterogeneous in their fitness and can keep each other balanced.
\end{abstract}
\maketitle

How biodiversity -- for example, the surprising coexistence of more than $10 \, 000$ species in a single gram of soil~\cite{Torsvik02, Daniel05, Gans05} -- is stabilised is one of the most fundamental questions in ecology~\cite{LevinMacArthurAward}. 
Known stabilising factors derived from resource competition models include metabolic trade-offs~\cite{PosfaiWingreen} and reciprocal oscillations in population sizes~\cite{HuismanWeissing}. 
Cyclic competition models and their derivatives are also frequently employed to model biodiversity \cite{Szolnoki,  XueGoldenfeldKillWinner, ShihGoldenfeld14}. 
However, recent experiments on a variety of different soil species in well-mixed pair-wise competition experiments found that these species could not be represented by cyclic competition models; instead, a few species outcompeted the others~\cite{HigginsGore}.
Thus, microbial diversity in soil is thought to be supported by the highly porous and fragmented structure of this habitat~\cite{YoungCrawford}, and the fluctuating environmental conditions that individual bacteria experience there~\cite{HigginsGore}.

Here, we ask how the influence of spatial structure, characterized by intrinsic variation between local environments, and the delayed adjustment to changes of these environments, can affect the long-term behaviour of species in a simple model system.
Indeed, it is well-known that such delays in physiological responses (here, adjustment of fitness or growth rate) occur in microbes, following externally imposed changes in the environment~\cite{AckermannReview, Lachmann, ThattaiOudenaarden, KussellLeibler}, such as  nutrient composition, or antibiotic stress~\cite{Monod49, LinKussell, FridmanBalaban, Bollenbach2009, YurtsevGore, WuAustin, SkanataKussell}. 
Here, we focus on changes that occur because species move between different habitats.  
We consider a system with two species, in which the dominant strain (fast grower) depends on the slower growing strain for its fitness. 
Thus, the fitness of both species depends on the intrinsic population structure of the local habitat. 
We assume that the fitness of an individual is not instantaneously reset when the environment changes, but is initially retained from its  previous environment, as we discuss below. 
\begin{figure}[t]
	\includegraphics[width=\columnwidth]{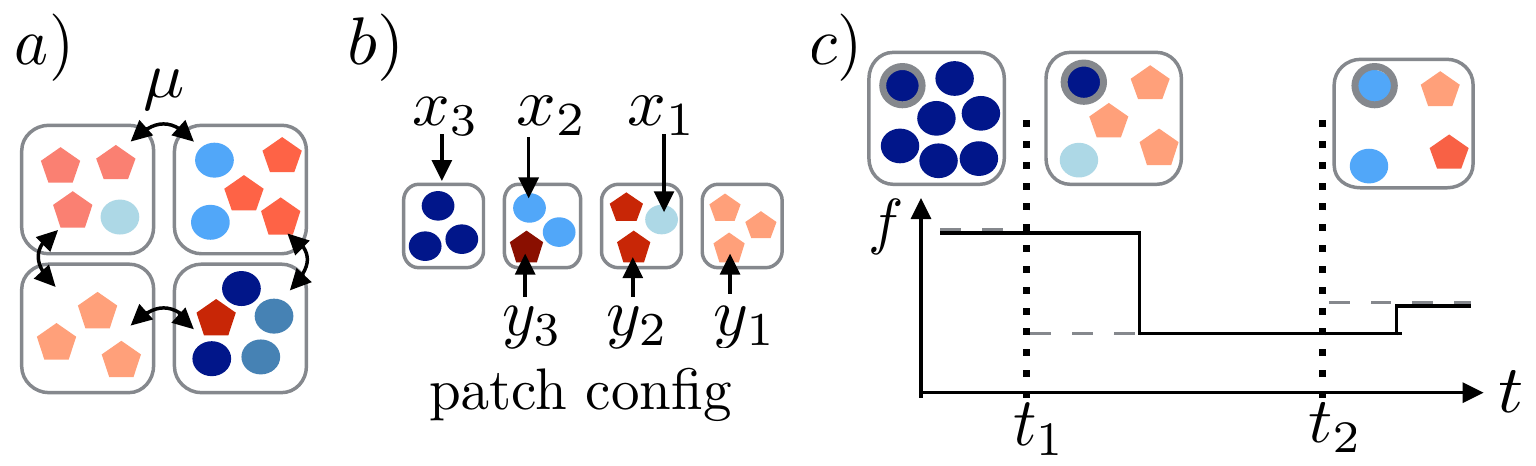}
	\caption{ (a) A slow-growing `producer' species (blue circles) and a fast-growing `non-producer' species (red pentagons) migrate in a metapopulation with mixing rate $\mu$; Shades indicate different fitnesses.
	(b) Reference fitness is defined by the species composition on a patch. 
	(c) The fitness of the circled individual (solid line) adjusts to the reference fitness (grey dashed line) with a delay, given by an adjustment rate $\omega$.  
 \label{fig:1}}
\end{figure}

We show that coexistence can arise in such a minimal two-species model, as a result of delays in fitness adjustment after a change of local habitat, and nonlinearity of fitness functions.
 In particular, we conclude that the combination of underlying spatial structure -- not spatial segregation -- and delay in adapting to environmental change can support biodiversity, by enabling individuals of the same species that have different fitness values to keep each other balanced.  
 A key result of our study is that, because of this delay in fitness change, the outcomes of direct competition experiments between species in well-mixed systems may differ from those observed in spatially structured habitats.
 
\emph{Model.}
We study a metapopulation of locally well-mixed patches. 
Individuals can move or hop between patches with rate $\mu$, which we refer to as the mixing rate, representing processes which couple individual patches (Fig.~1a). 

The local species composition on a patch $\nu$ with $N_\nu$ individuals determines their reference fitness on that patch. 
A well-known implementation of such an eco-evolutionary model is, for example, a public goods dilemma~\cite{Velicer2003}. 
There, a `producer'-species produces a public good, which bestows a fitness benefit on the entire population~\cite{RaineyRainey03, GreigTravisano, WestGriffinGardnerDiggle, diggle07, GoreOudenaarden, CorderoPolz12, Drescher2014, Nadell16, Becker17}, while imposing a fitness cost on its producer. 
Hence, the slower growing producer is at a disadvantage relative to a faster growing  `non-producer' species~\cite{Hardin1968, Axelrod1981}.
The local amount of public good (such as invertase for yeast or glutamine synthetase for \emph{B.~subtilis}~\cite{GoreOudenaarden, LiuSuel}) on a given patch increases in general with increasing producer fraction on that patch, which in turn enhances the fitness of both species on the patch. The effect on the fitness can vary nonlinearly: the impact  of the public good on an individual's fitness might saturate when a large amount of public good is present (see e.g.~Refs.~\cite{GoreOudenaarden, ArchettiFerraroChristofori, HeilmannKerr}), or only be noticeable when a significant number of producers is present on the patch~\cite{Cornforth12}.

We study species that interact via such public goods as an exemplary interaction topology that results in hierarchical population structure. 
As the precise dependence of a species' fitness on species composition can vary between different experimental setups~\cite{DamoreGore}, it makes sense to study a conceptual model, which minimizes the number of tunable variables. 
Here, we assume that the growth rate or fitness of individuals depends on the fraction of slow growers on a patch, $N^{-}_\nu/N_\nu$,
\begin{equation} 
\label{eq:F}
	f_{\nu} = f_0 + \big ( N^{-}_\nu /N_\nu \big) ^\alpha \textrm{,}
\end{equation}
where we set the basal level of fitness $f_0 \,{=}\, 1$ to ensure that the fast grower can grow in absence of the public good. A nonlinearity with $\alpha\,{<}\,1$ implies that the public good  provides less additional benefit for high numbers of slow growers.  
We distinguish between the fast-growing and slow-growing strain by introducing a fitness difference $c$ between them, which measures how much slower the slow-grower would reproduce in the exact same environment.
Thus, the reference fitness of the fast and slow growers in the same local environment are $f_{\nu}^{+} \,{=}\, f_\nu$  and $f_{\nu}^{-} \,{=}\, f_{\nu}-c$, respectively.  We note that this type of interaction, where the fitness depends on a group of $N_\nu$ players, is referred to as an $N$-player game  in the context of game theory~\cite{MitteldorfWilson, Taylor92, vanBaalenRand, ArchettiScheuring}.

Because of the different numbers of slow growers that can inhabit a patch, individuals of the same species may have different fitness values (Fig.~1). 
For a local patch with $N$ individuals, there are $N \,{+}\, 1$ different possible compositions and thus $N$ different \emph{fitness types} for each species (Fig.~1b). 
These different fitness types have fitness values $f^{-}_{i} \,{=}\, f_0 \,{+}\, \big (i /N \big) ^\alpha \,{-}\, c$ for $i \in [1,N]$ for the slow growers and $f^{+}_{i+1} \,{=}\, f_0 \,{+}\, \big (i /N \big) ^\alpha$ for $i \in [0,N \,{-}\, 1]$ for the fast growers. Here, $i$ stands for the numbers of slow growers on the patch, and  we have shifted the index so that the fitness of the least fit individual can be denoted by $f_{1}^{\pm}$ in both cases.
Since the number of individuals may differ between patches, there is a large variety of different fitness types in the  metapopulation at any one time.

The local species composition thus determines the reference fitness of an individual, which in turn determines its growth rate: 
Individuals reproduce at rates proportional to their fitness values by replacing another individual on the same patch by an identical copy of itself. 
This implementation of reproduction via the so-called Moran process~\cite{Moran62} is a technicality and ensures that the number of individuals on a patch does not change, thereby avoiding stabilisation of the slow growers by known effects which we do not consider here, such as by disproportional growth of slow grower patches~\cite{SimpsonChuang2009, Jonas_and_anna}. 

\emph{Delay.}
So far, we have specifically used the term `reference fitness' in order to distinguish between the fitness prescribed by the local environment, and the actual fitness value of an individual.
We assume that there is a delay between the change in species composition of a patch and the time at which the individual adjusts to the reference fitness (see Fig.~1c for the fitness of an exemplary slow grower in a short span of its lifetime). 
This assumption seems appropriate, as the exoproduct concentration on the patch will require some time to equilibrate, due to, for example, slow production of exoproducts,  slow and heterogeneous detection of changes in patch composition~\cite{WolfBischofs, SnijderPelkmans, RaserOShea}, or  slow local diffusion of the produced public good~\cite{AllenGoreNowak, BorensteinShaevitzWingreen}.  
We absorb all these  biological processes into one stochastic fitness adjustment rate $\omega$ for each individual. 
Thus, Fig.~1c shows that the slow grower retains its fitness from the previous environment for a period that scales on average as $1/\omega$.

\emph{Coexistence.}
We first discuss how the presence of a delay in fitness adjustment alters the species composition of the entire metapopulation. 
Without this delay ($\omega \,{\to}\, \infty$), the slow grower would consistently die out for all parameters if the fitness difference is finite, $c>0$: it dies out because it reproduces at a slower rate than the fast grower,   
 and because we choose the system size $L$ large enough for stochastic fluctuations not to influence the outcome~\cite{Shnerb2000, Traulsen05, Traulsen06, Reichenbach2006,  Mauroreview}.

We explore the effect of delay with a stochastic Gillespie simulation~\cite{Gillespie} of individuals which get mixed between patches with rate $\mu$, adjust their fitness with rate $\omega$, and reproduce with their actual fitness, on a two-dimensional square lattice of $L\,{\times}\,L$ patches. We show the phase diagram of the stability of the species for fitness functions with $\alpha\,{<}\,1$ in Fig.~2a: here we used the average extinction time to characterize different stability regions (as opposed to the population composition, cf.~Fig.~S4 in the supplement). 
Strikingly, a phase of coexistence of both species, marked by large coexistence times of $T_{\textrm{ext}}\,{>}\,7500$,  extends over a broad regime of fitness difference $c$ and mixing rates $\mu$. The coexistence phase begins at mixing rates less than the basal fitness $f_0\,{=}\,1$, and broadens with increasing mixing rate. 
As it is most pronounced for high mixing rates, we explain its presence first by discussing the well-mixed limit for all different values of $\alpha$.
In doing so, we will also show why coexistence only occurs for $\alpha\,{<}\,1$.
\begin{figure}[t]
        \includegraphics[width=\columnwidth]{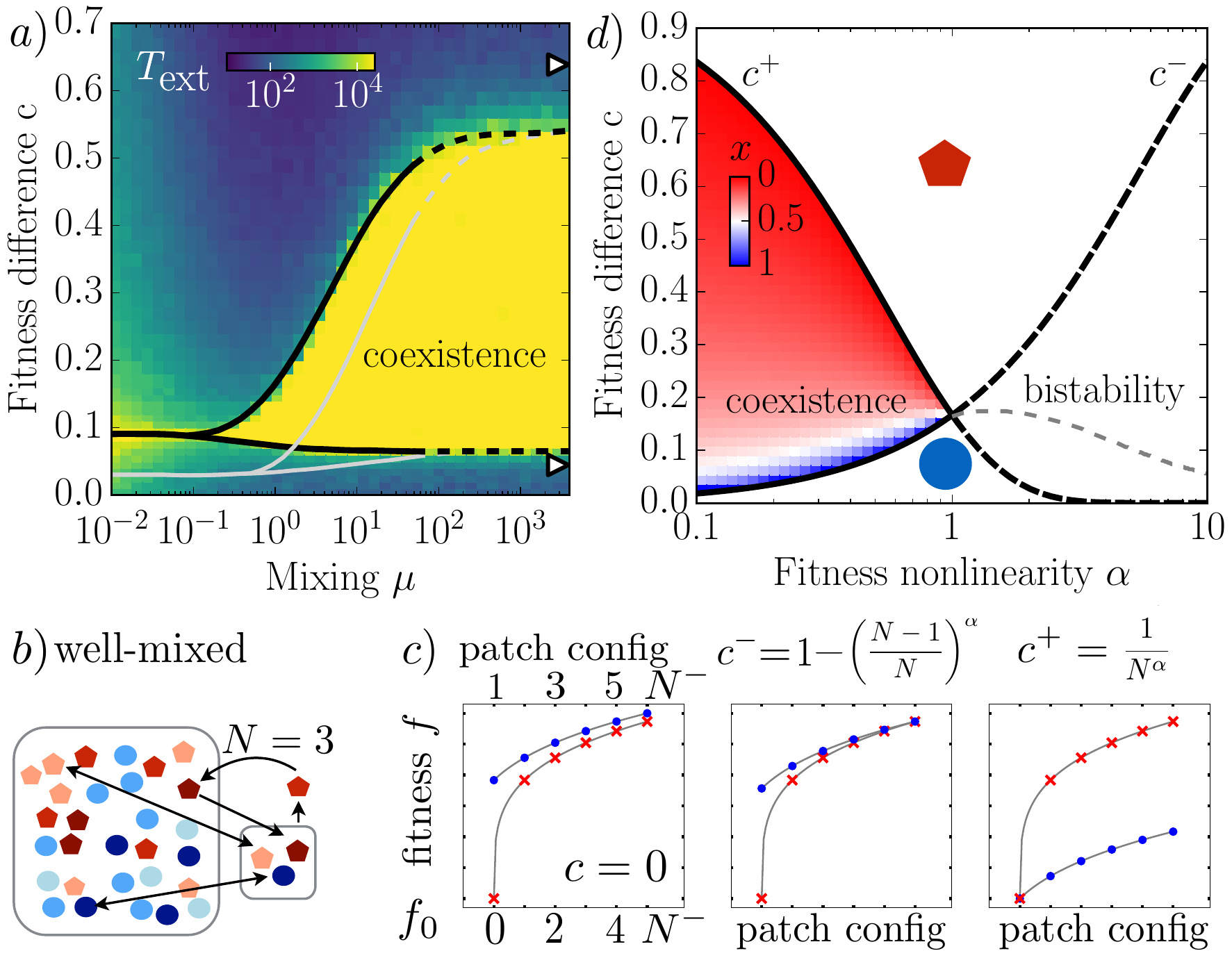}
\caption{Effect of fitness nonlinearity on species stability: 
	(a) Stability phase diagram, showing average extinction time  (ca.~$20$ simulation runs for $\omega \,{=}\, 5$, $\alpha \,{=}\, 0.25$, $L \,{=}\,30 \times 30$, initial $N \,{=}\, 6$). Coexistence occurs over a large range of $\mu$ and $c$. We take coexistence to be present when neither species becomes extinct prior to a  long cutoff time ($T_\textrm{ext} \,{>}\,7500$ for 5400 individuals). The boundaries of this region (lines) change only qualitatively for $\omega \ge 1$ (black vs grey line for $\omega \,{=}\, 20$, marked in dashes when  simulations with different system sizes suggest that finite size effects dominate).   
	(b) In the well-mixed limit, reproduction occurs across all patches, but fitness is defined by groups of $N$ individuals. 
	(c) Sketch of possible fitness values for $N\,{=}\,6$, $\alpha \,{=}\, 0.25$: coexistence can arise for $c^{+} \,{:=}\,1/N^\alpha \,{<}\, c \,{<}\, 1 \,{-}\, ((N{-}1)/N)^\alpha\,{=:}\,  c^{-}$.  
	(d) The fixed point structure of the well-mixed limit (for $N \,{=}\, 6$) shows that species coexist for sizeable ranges of fitness difference between the solid lines for $\alpha \,{<}\, 1$, and that the system is bistable between the dashed lines for $\alpha \,{>}\, 1$. (Colorscale: slow grower fraction at this fixed point). The cost up to which the slow grower survives for initial conditions of equal fast and slow growers for $\alpha \,{>}\, 1$ is shown with the grey dashed line.
\label{f:inv}}
\end{figure}

\emph{Well-mixed limit.}
The delay in attaining the reference fitness means that different fitness types within a species (with fitnesses adjusted to previous environments) are present on a local patch in the well-mixed limit. 
Reproduction then occurs within a random sample of the fitness types present in the well-mixed metapopulation, 
which is equivalent to reproduction occuring across the entire population with all different fitness types.

On average, the system can then be described by a set of differential equations, one for each fitness type (see section `Structure of Differential Equations' in the supplement). 
We consider the average number of $\overline{N_\nu}=N$ individuals in the following, even though the number of individuals per patch may vary in principle due to mixing. 
Doing so means that there are then only $N$ distinct fitness types per species in the metapopulation (Fig.~1b). 
 In this case, the ensuing mean-field dynamics is given by a set of $2 N$ coupled differential equations for the density of each fitness type: We denote the density of a slow grower of type $i \,{\in}\, [1,N]$ by $x_i$ with $x \,{=}\, \sum_i x_i$ (Fig.~2b); the same applies for fast growers ($y$).
Each differential equation contains a reproduction term, to describe the Moran process, and a term for the fitness adjustment, which occurs with rate $\omega$. In the stochastic model, fitness adjustment  involves the species composition at the patch of the adjusting player, yet in the well-mixed limit  this species composition corresponds to a randomly drawn sample of individuals from the metapopulation. For  large numbers of individuals in the system, $L\,{\times}\,L\,{\gg}\,N$, randomly drawing a sample of individuals turns into a combinatorial problem involving the densities in the metapopulation (see supplement). 
 The time evolution for the $i^{\textrm{th}}$ slow grower fitness type of density $x_i$ is then described by
\begin{align} \label{eq:diffeq}
	&	\frac{dx_i}{dt} {=}\bigg( f^{-}_{i} x_i (1-x_i) {-} \sum_{j} f^{+}_{j} y_j x_i {-} \sum_{j\ne i} f^{-}_{j} x_j x_i \bigg)   \\
	& \quad{-} \omega \bigg( \sum_{j\neq i}^{} j \binom{N}{j} x_i x^{j-1} y^{N -j} {-} i \binom{N}{i} ( x-x_i )x^{i-1} y^{N -i}\bigg) \textrm{,} \nonumber
\end{align}
where  the sum over $j$ runs over all fitness types unless otherwise specified. 

Before analyzing the fixed point structure of Eq.~\ref{eq:diffeq}, we use arguments from game theory to explain  how coexistence can arise~\cite{MaynardSmith, DrosselAdv, NowakSigmund04, FreyReview}. 
To do so, we investigate the evolutionary stability of the absorbing states (all fast or all slow growers) by comparing the fitness of individuals in these absorbing states to the fitness of individuals in a state with a single individual of the other species. To illustrate this, the panels in Fig.~2c show the fitness of slow and fast growers for all possible patch configurations 
(shown in Fig.~S1  or for $N \,{=}\, 3$ in Fig.~1b), with increasing numbers of slow growers. 
The configuration of the fittest slow-grower contains $N$ slow growers, while the configuration for the fittest fast-grower contains only $N {-} 1$ slow growers. 
Hence, for zero fitness difference $c\,{=}\,0$, the state of all slow growers is the fittest state overall (left panel Fig.~2c).
Thus for $c \,{=}\, 0$, the absorbing state of all slow growers is evolutionarily stable.

An increasing fitness difference reduces the fitness  of the slow grower. 
The fitness of the slow grower in the absorbing state and the fittest fast grower are equal at $c^{-} \,{=}\, 1{-}((N{-}1)/N)^\alpha$ (Fig.~2c, middle panel). 
Hence the state of all slow growers becomes invasible for $c \,{>}\, c^{-}$, and is no longer evolutionarily stable. However, the state of all fast growers is also not evolutionarily stable, because a single slow grower surrounded by  fast growers is fitter than the fast growers in the absorbing state. As neither absorbing state is stable, an intermediate fixed point  at which both species coexist must be stable. This coexistence fixed point is stable up to $c\,{=}\,c^{+}\,{=}\,1/N^\alpha$, where the state of all fast growers becomes evolutionarily stable.

This game theoretical analysis is corroborated by a linear stability analysis of the fixed points of the set of Eq.~\ref{eq:diffeq}.
As the Jacobian evaluated at the absorbing fixed points is sparse (cf. supplement), its eigenvalues are easily accessible. 
For $\omega\,{>}\,f_0$ as considered here, all eigenvalues apart from one are always negative. 
The eigenvalue that can change sign is given by $f^{-}_{1} \,{-}\, f^{+}_{1} \,{=}\,c^{+}-c$ for the fast, and $f^{+}_{{N}} \,{-}\, f^{-}_{N}\,{=}\,-c^{-}+c$ for the slow grower absorbing state. 
Thus, the fast (slow) grower fixed point is stable for $c \,{>}\, c^{+}$ ($c \,{<}\, c^{-}$). 

We plot $c^{-}$ and $c^{+}$ and thus the regions of fixed point stabilities in Fig.~2d for $N \,{=}\, 6$. 
Since for $c\,{<}\,c^{-}$ slow growers are stable (and vice versa) it is clear from Fig.~2d that coexistence occurs for $\alpha\,{<}\,1$, where $c^{-}\,{<}\,c\,{<}\, c^{+}$ is possible. Conversely, for $\alpha \,{>}\, 1$ the system is bistable for $c^{+} \,{<}\, c \,{<}\, c^{-}$. For better intuition of how the system would behave for $\alpha \,{>}\, 1$, we show $c$ up to which the slow grower survives for nonbiased initial conditions  with the grey dashed line. The shading  for $\alpha \,{<}\, 1$ shows that the  fraction of slow growers at the coexistence fixed point (obtained from numerical solution of Eq.~\ref{eq:diffeq}, cf.~supplement) increases with fitness difference $c$.

This fixed point structure is robust:  
we discuss in the supplement how the extent of the coexistence region depends on $N$ and $\alpha$.
Importantly, the region of coexistence in the well-mixed limit does not depend on the length of delay unless $\omega \,{<}\, f_0$, as the eigenvalues retain the structure discussed above for $\omega \,{>}\, f_0$ (for details on the effect of $\omega$ on coexistence, see Ref.~\cite{BauerFreyEPL}). 
This condition on $\omega$ ensures that the fitness adjustment is slow enough for different fitness types to be generated repeatedly between reproduction events.

Coexistence can arise here because the fittest slow growing individual can be fitter than the fittest fast grower. This is possible due to the combination of $N$-player interaction and nonlinear fitness function~\cite{Motro, Hauert06, ArchettiScheuringEvolution11, AssafMobiliaRoberts, ArchettiScheuring, LiThirumalai}. We emphasise that this combination alone does not lead to coexistence in our model: only the additional timescale~\cite{Levin2000, GoldenfeldWoese, RocaSanchez09, WuHolme} due to delay introduces phenotypes which differ in their fitness, and can keep each other in balance. This mechanism for coexistence also applies to larger groups of microbes: a fittest slow grower which temporarily maintains its fitness  due to delay can occur in a variety of microbial interactions, for example in models including private public goods~\cite{Jung2018, LiThirumalai, GoreOudenaarden, ZhangRainey}. 

 Strikingly, coexistence not only arises in the well-mixed limit, but also for a wide parameter range of more realistic mixing rates: Fig.~2a shows that even when the same percent of the population is mixed as can reproduce per unit time, coexistence can occur. That coexistence extends to parameters where the system is no longer well-mixed is  not clear a priori. The physical intuition about the balance of fitness types from the well-mixed limit is thus helpful to understand the phenomenology in the complex, spatially extended system.

\emph{Phase diagram.}
The phase diagram in Fig.~2a shows the average extinction times and phase boundaries for $\omega \,{=}\, 5$, and a shorter delay of $\omega \,{=}\, 20$ for comparison for nonlinear fitness functions with $\alpha \,{=}\, 0.25 $. We verified that the simulations that were terminated after $T \,{>}\, 25\, 000$ yield slow- and fast-grower coexistence at the end of their run, and show the slow grower density for exemplary simulation runs for all three phases in Fig.~S5. This fixed point density from stochastic simulations for finite, yet high mixing rates oscillates about values close to the deterministic well-mixed limit, with slight deviations due to correlations introduced by the slower mixing rate, and the fact that the number of individuals per site varies in the stochastic simulations. 

The white triangles in Fig.~2a mark the fitness differences $c$ between which coexistence is expected from the well-mixed limit (cf.~Fig.~2d).  
Our phase boundaries do not exactly match the well-mixed expectation, as stochastic fluctuations in a finite size system can lead to spontaneous extinction of one species, especially when the density of that species is expected to be small at the stable fixed point. We verified the phase boundaries by changing the size $L$ of the metapopulation, and marked the phase boundaries with dashes where variations in $L$ suggested that system size effects to dominate.

In the weak mixing limit, the phase boundary marks a transition from a fast growing to a slow growing phase without a coexistence region. 
The phase boundary in this limit can be determined by considering local patches on which one of the species has established, and then comparing the probabilities of single individuals to invade another patch~\cite{AltrockTraulsen, AntalScheuring, BlytheMcKane, BauerFrey}. Because a single fitness type fixes when only one species is present on a patch, coexistence --which relies on multiple fitness types-- is not possible for very low mixing rates.
For $\alpha \,{=}\, 1$, the transition between the low and high mixing limit occurs approximately at $\omega \,{\approx}\, \mu$~\cite{BauerFrey}. 
For $\alpha \,{<}\, 1$, the two phase boundaries in Fig.~2a indicate that the lowest mixing rate at which the coexistence occurs also increases with $\omega$; 
we verified this general trend, but a precise investigation of these values would go beyond the scope of this work, as the phase diagram remains qualitatively the same.

We find it remarkable that coexistence can occur over such a wide parameter range, as this feature indicates that the stabilisation of populations afforded by delayed fitness adjustment  may be a realistic effect. 
We note that our results \emph{do not} depend on the formation of spatial patterns~\cite{Nowak92, Reichenbach2007, PigolottiMunoz}.

\emph{Conclusion.}
We have shown that species can coexist in spatially fragmented systems with delays in fitness adjustment even when one species always dominates over the other in a locally well-mixed environment.
This coexistence occurs because the combination of delay with the interaction via a public good means that species compete via a variety of different fitness types (cf.~Fig.~2a) which can balance each other for some fitness functions. 
While this conceptual work studies a general  fragmented environment and does not specifically model soil (see e.g. Refs.~\cite{Coyte2017, PeaudecerfCroze, PerezReche, Dechesne} for important aspects of the soil structure), the robustness of coexistence over large parameter ranges suggests that delayed responses to changes of  habitat may be a significant factor in the maintenance of biodiversity.
Thus, our study indicates that it is important to mimic the spatial structure when biodiversity in spatially structured habitats is investigated (see e.g. Refs.~\cite{vanDyken13, Tarnita15, GandhiKorolevGore,PandeKost16,PeaudecerfCroze,  AlfonsoGore}), even in experimental configurations where spatial assortment  does not occur. 

\emph{Acknowledgement.}
 We are grateful to Jonas Denk, Nigel Goldenfeld,  Heinrich Jung, Felix Kempf and David Muramatsu for discussions, and acknowledge funding from the European Union Horizon 2020 research and innovation programme under the Marie Sk\l{}odowska-Curie grant agreement No 660363, an LMU Research Fellowship, and the Volkswagen Foundation via the programme `Life? - A Fresh Scientific Approach to the Basic Principles of Life'.

\pagebreak
\newpage
$\phantom{i}$ \newpage

\widetext
\begin{center}
\end{center}
\setcounter{equation}{0}
\setcounter{figure}{0}
\setcounter{table}{0}
\setcounter{page}{1}
\makeatletter
\renewcommand{\theequation}{S\arabic{equation}}
\renewcommand{\thefigure}{S\arabic{figure}}
\renewcommand{\bibnumfmt}[1]{[S#1]}

\section{Supplemental Information}
\section{Structure of differential equations in the well-mixed limit}

In this section, we discuss the structure of the differential equations describing the well-mixed limit. Because the mean dynamics of the system are sufficient to capture the phenomenology, as we show in the following, we do not analyse the more complex stochastic Master equation.

In the well-mixed limit, the mean dynamics of the system should in principle be described by a replicator-type equation, in which individuals reproduce with their actual fitness. For a general system in which different individuals of two species have different fitness values, this replicator equation involves a sum over the total possible number $J$ of different fitness values for the slow growers and over the total number $K$ of different fitness values for the fast growers (which need not necessarily be the same at any one time). If we define the population fraction of slow growing individuals with fitness  $f^{-,j}$ as $x_{j}$ and of fast growing individuals with fitness $f^{+,k}$ as $y_{k}$, this general replicator equation reads
\begin{align} \label{eq:11}
	\frac{d x}{dt}  & = \sum_{j=1}^{J} f^{-,j} x_j (1-x)  - \sum_{k=1}^{K} f^{+,k} y_k x \textrm{,}
\end{align}
We placed the indices as a superscript on the fitness values $f^{-,j}$ and $f^{+,k}$, in order to avoid confusion with the different fitness types for an average of $N$ individuals per patch (as discussed in the main text and below). As in the main text, $x$ denotes the density of slow growers and $y=1-x$ the density of fast growers.

The different possible fitness values in Eqn. \ref{eq:11} can reflect the population composition on a patch at a previous time, which may be different for every individual. Thus, one may initially think that the number of different possible fitness values $J$ and $K$ are very large. However, each individual is equally likely to be selected for fitness adjustment (with rate $\omega$). Once an individual is selected, the fitness to which its fitness is adjusted depends only on the current species composition on its patch, and not on its history. (We note that this means that the dynamics of the system here is structurally simpler than that for typical delay differential equations, where the new state depends explicitly on the state of the system at a previous time, see e.g. Refs.~\cite{Gopalsamy,Faria,Chen}.)

Thus, there are only as many different fitness types in our system as there are different patch configurations, i.e. different species compositions on the patch. In principle, the number of individuals per patch can vary when individuals move between patches. We simplify the system by considering an average number of $N\,{=}\,\overline{N_\nu}$ individuals on a patch. In this case, the possible fitness values at any point in time are significantly restricted, and can only take up one of a set number of values, or fitness types:  For $N \,{=}\, 6$, which we discuss in this supplement, we then need to consider $N  \,{=}\,  6$ different fitness types of slow growers with fraction  $x_i$ for $i \in \{1, 6\}$, and $N  \,{=}\,  6$ different types of fast growers with total population fraction  $ y_i$ for $i\in \{1,6\}$ (corresponding to $N^{-}\, {\in}\, \{1,N\}$ and $N^{-}\, {\in}\, \{0,N-1\}$ slow growers in the local environment, respectively).

\begin{figure}[t]
        \includegraphics[width=0.35\columnwidth]{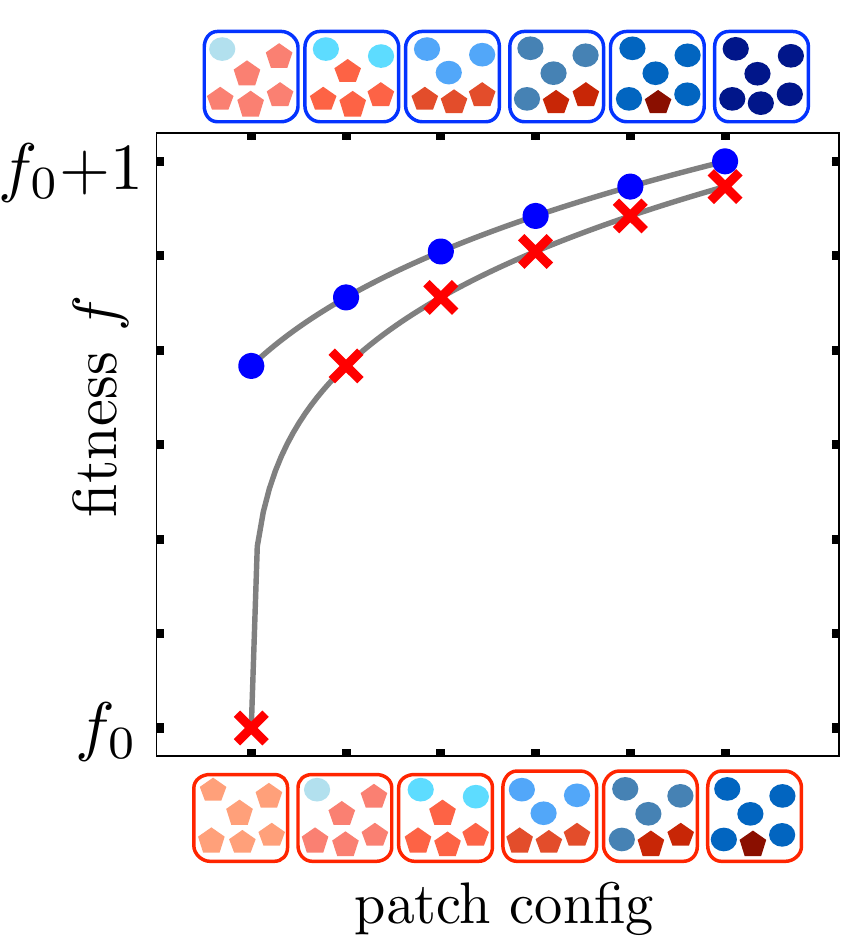}
	\caption{Enlarged Fig.~2c (left panel) of the main text, with the different patch configurations showing the population composition from weakest to strongest fitness type of each species: reference fitness values of the different slow growers (top $x$-axis) and fast growers (bottom $x$-axis) in the different possible environments for $N \,{=}\, 6$ players. \label{f:supp2}}
\end{figure}

Figure~\ref{f:supp2} shows the fitness values of all different fitness types. This figure is a reproduction of the left panel of Fig.~2c in the main text, with the only difference that here we show the full patch configuration representing the different fitness types for both species on the top and bottom $x$-axis (slow and fast grower, respectively), for clarity.

Because these fitness types are discrete, and because fitness adjustments can be interpreted as flow between these different fitness types, it is easier to think about a set of differential equations for the densities $x_i$ of individuals of each possible fitness values $f_{i}^{-}$ or $f_{i}^{+}$ in the entire system, instead of one differential equation for $x$. Each of these differential equations is made up of a reproduction term, $d x_{i,\textrm{repro}}/dt$, and a term for fitness adjustments, $d x_{i,\textrm{update}}/dt$. This latter term for fitness adjustments contains the rates with which a particular fitness type is generated during a fitness adjustment event, which depends on the species composition on the patch at which the adjustment is performed.

The temporal change $d x_{i}/dt$ for each fitness type $x_i$, can then be split into a term corresponding to reproduction, $d x_{i,\textrm{repro}}/dt$, and a term for fitness adjustments, $d x_{i,\textrm{update}}/dt$, which we explain separately in the following.

Reproduction can occur between any two individuals in the well-mixed limit and occurs proportionally to an individual's fitness, such that part of the differential equation for reproduction, $d x_{i,\textrm{repro}}/dt$, reads
\begin{equation*}
	\frac{d x_{i,\textrm{repro}}}{dt} = f_{i}^{-} x_i (1-x_i) - \sum_{j=1}^{N} f^{+}_{j} y_j x_j - \sum_{j=1, j \neq i}^{N} f^{-}_{j} x_j x_i \textrm{.}
\end{equation*}
The first term corresponds to individuals of type $x_i$ reproducing by replacing any other individual in the well-mixed metapopulation, while the other two terms symbolise the reproduction of another species by replacement of $x_i$.
This type of reproduction means that $x+y \,{=}\, 1$, i.e. the number of individuals in the metapopulation is conserved.

Fitness adjustments only occur within a species (i.e. a slow grower can adjust its fitness to a different value, but will still remain a slow grower). Each individual adjusts its fitness on a patch, or in the well-mixed limit discussed here, in a group of $N$ individuals. The rate with which an individual becomes an individual of a certain fitness type thus depends on the species composition on its patch. Since one assumes that there are no correlations between patches in the well-mixed limit, and hence that the species composition on a patch corresponds to the species composition of a randomly drawn sample from the metapopulation. We assume that the metapopulation so large that that total number of individuals in the entire metapopulation is much larger than the number of individuals in a group, $L\,{\times}\, L\,{\times}\, N\,{ \gg}\, N$ (or $L\,{\times}\, L\,{ \gg}\,1$). In addition, we assume that this metapopulation also is so large that many different (uncorrelated) groups can be formed, $L\,{\times}\, L\,{ \gg}\, N$. In this case, the probability to find a certain species composition on any one patch (and thus the rate to become a certain fitness type) can be calculated combinatorically from the densities of two species in the entire metapopulation.
 Since only the number of slow growers matters for defining the new fitness type in a fitness adjustment event, we need to calculate the mean probability that there are $N^{-}$ slow growers in this group of $N$ individuals, given the slow grower density in the metapopulation, $x$.  For example, the fraction of individuals of the least fit slow-growing type of fitness $f_0 + \big ( 1/N \big)^\alpha-c$ increases when fitness adjustments of a slow grower occur in groups that contain precisely one slow grower: for this term, one thus has to consider all different combinations of individuals with only one slow grower and five fast growers, i.e. $(x_2+x_3+x_4+x_5+x_6) y^5$. The term is weighed by all different possibilities with which this combination can be drawn out of the entire metapopulation, which in this case adds up to ${6}\choose{1}$. Similarly, there are fitness adjustment events which lead to a decrease in the fraction of this fitness type, for example when this least fit slow grower has adjusts to an environment with one further slow grower (of any type) and four fast growers. All other terms can be calculated analogously.
Thus, for an average of $N \,{=}\, 6$ individuals per patch, the differential equations for fitness adjustment (here referred to as $d x_{i,\textrm{update}}/dt$) for the slow growers read
\begin{widetext}
\begin{align*}
        \frac{d x_{1,\textrm{update}}}{dt} &=  {{6}\choose{1}}(x_2+x_3+x_4+x_5+x_6)y^5 -2{{6}\choose{2}}x_1 x y^4 - 3{{6}\choose{3}}x_1 x^2 y^3 - 4 {{6}\choose{4}}x_1 x^3 y^2 - 5 {{6}\choose{5}} x_1 x^4 y - 6 {{6}\choose{6}} x_1 x^5 \textrm{,}\\
        \frac{d x_{2,\textrm{update}}}{dt} &= - {{6}\choose{1}}x_2 y^5 + 2{{6}\choose{2}}(x_1 +x_3+x_4+x_5+x_6) x y^4 - 3{{6}\choose{3}}x_1 x^2 y^3 - 4 {{6}\choose{4}}x_2 x^3 y^2 - 5 {{6}\choose{5}} x_2 x^4 y - 6 {{6}\choose{6}} x_2 x^5 \textrm{,}\\
        \frac{d x_{3,\textrm{update}}}{dt} &= - {{6}\choose{1}}x_3 y^5 -2{{6}\choose{2}}x_3 x y^4 + 3{{6}\choose{3}}(x_1 +x_2+x_4+x_5+x_6) x^2 y^3 - 4 {{6}\choose{4}}x_3 x^3 y^2 - 5 {{6}\choose{5}} x_3 x^4 y - 6 {{6}\choose{6}} x_3 x^5 \textrm{,}\\
        \frac{d x_{4,\textrm{update}}}{dt} &= - {{6}\choose{1}}x_4 y^5 -2{{6}\choose{2}}x_4 x y^4 - 3{{6}\choose{3}}x_4 x^2 y^3 + 4 {{6}\choose{4}} (x_1 +x_2+x_3+x_5+x_6) x^3 y^2 - 5 {{6}\choose{5}} x_4 x^4 y - 6 {{6}\choose{6}} x_4 x^5 \textrm{,}\\
        \frac{d x_{5,\textrm{update}}}{dt} &=  -{{6}\choose{1}} x_5 y^5 -2{{6}\choose{2}} x_5 x y^4 - 3{{6}\choose{3}}  x_5 x^2 y^3 - 4 {{6}\choose{4}} x_5 x^3 y^2 + 5 {{6}\choose{5}}( x_1+x_2+x_3+x_4+x_6) x^4 y - 6 {{6}\choose{6}} x_5 x^5 \textrm{,}\\
        \frac{d x_{6,\textrm{update}}}{dt} &=  -{{6}\choose{1}} x_6 y^5 -2{{6}\choose{2}} x_6 x y^4 - 3{{6}\choose{3}}x_6 x^2 y^3 - 4 {{6}\choose{4}} x_6 x^3 y^2 + 5 {{6}\choose{5}} x_6 x^4 y + 6 {{6}\choose{6}}(x_1+x_2+x_3+x_4+x_5) x^5  \textrm{.}
\end{align*}
\end{widetext}
These terms need to be multiplied by fitness adjustment rate $\omega$. As discussed for the least fit slow grower, there is always one positive term for each $x_i$ (the term with $i$ slow growers is positive), since  fitness adjustment in this group would increase the number of $x_i$ in the system, while all other terms are negative. The binomial term in front of each term indicates how many possibilities there are to distribute the number of fast growers among the $N$ individuals, while the number before the binomial counts the possibilities to order the specific slow grower among the other slow growers.

The full differential equation for any slow grower is then 
\begin{align} \label{eq:diffeq2}
	\frac{d x_{i}}{dt}= \frac{d x_{i,\textrm{repro}}}{dt}+ \omega \frac{d x_{1,\textrm{update}}}{dt} \, .
\end{align}
The concise form for these differential equations for {general} $N$ is shown in the main text of the paper.
The differential equations for different fitness fractions $y_i$ for the fast growers can be expressed analogously. 
The constraint $\sum_i x_i + y_i  \,{=}\, 1$ can be used to eliminate one of these $2N$ differential equations.

It is clear that $x \,{=}\, 0$ (more precisely $y_1 \,{=}\, 1$) and $x \,{=}\, 1$ (more precisely $x_{N} \,{=}\, 1$) are both fixed points of the systems, corresponding to survival of the weakest fast and strongest slow grower. We next explore whether these fixed points are stable for general $N$.

\begin{figure}
\includegraphics[width=0.9\columnwidth]{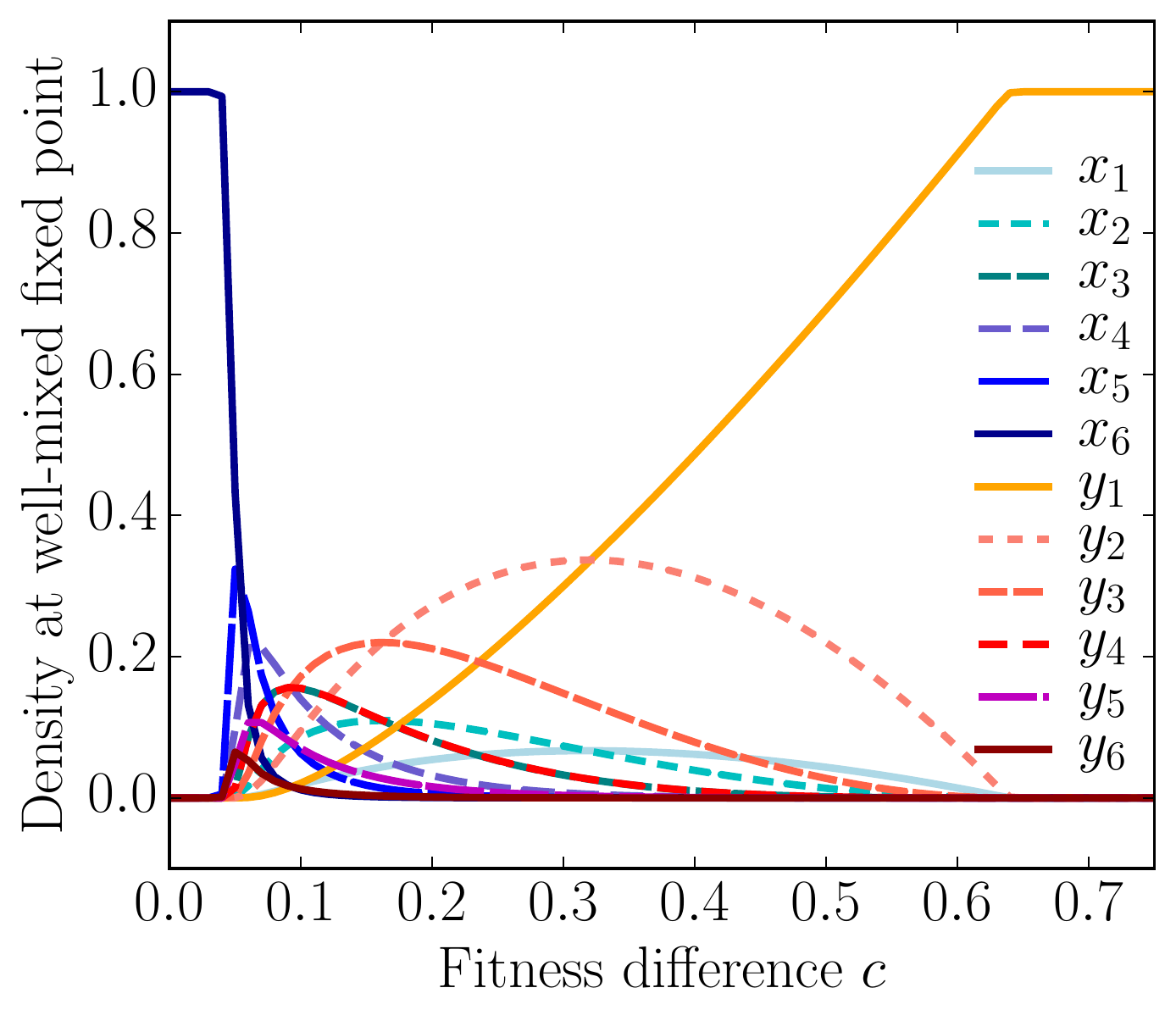}
\caption{ 	
	Species composition at coexistence fixed point from the deterministic differential equations from numerical solution main text eq.~2 or eq.~\ref{eq:diffeq2} (same parameters as Fig.~2a in the main text), from numerical solution. Slow grower (fast grower) densities are shown in blue (red). Between $c^{-}$ and $c^{+}$ (for definition see text), a variety of fitness types are present and keep each other balance, while for $c\,{<}\,c^{-}$ ($c\,{>}\,c^{+}$) only the slow or fast growers are stable.
\label{f:fixedpoint}}
\end{figure}

\begin{figure}
	\includegraphics[width=0.9\columnwidth]{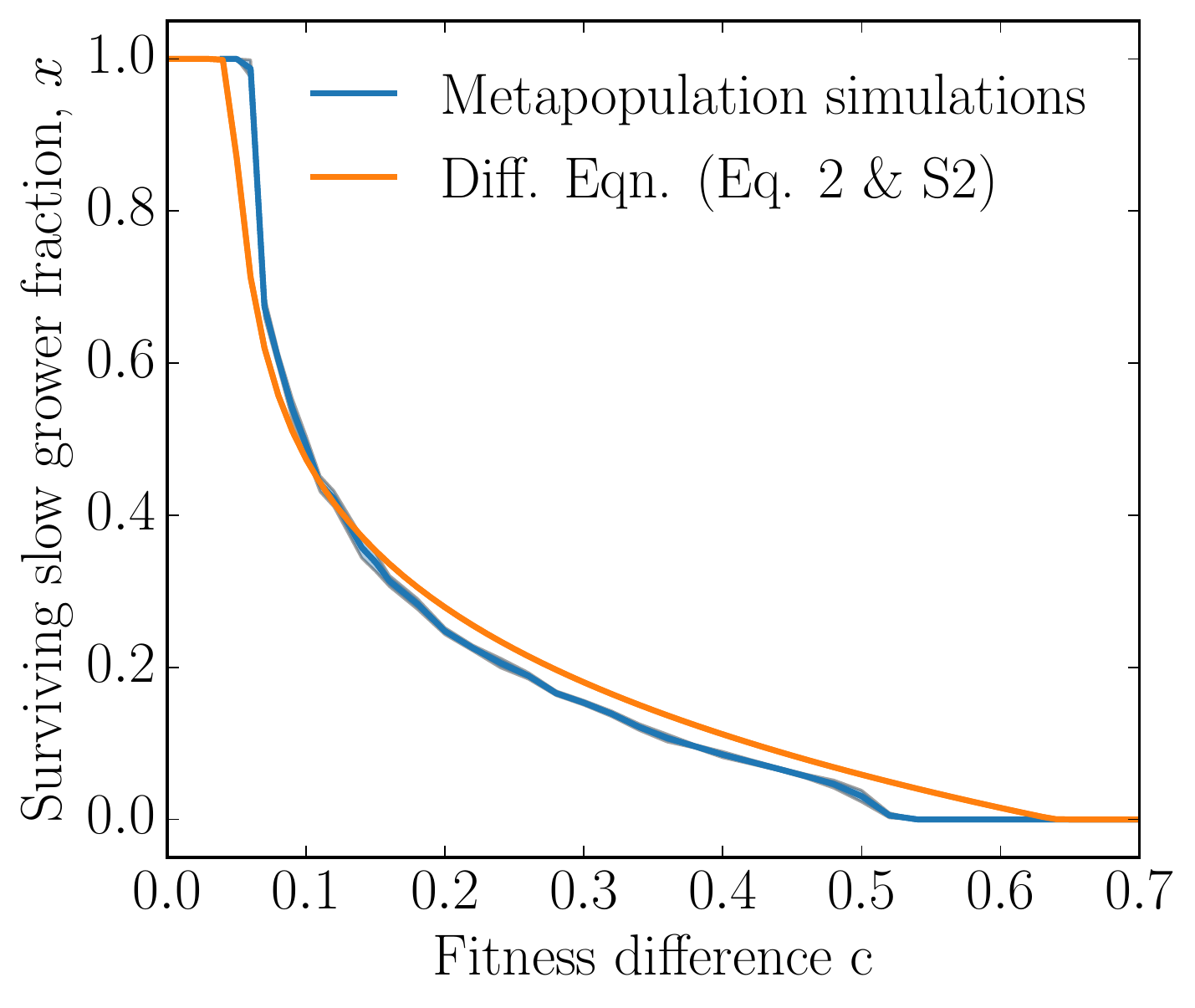}
\caption{ 	
Comparison of the slow grower fraction at the end of the stochastic Gillespie simulations for the metapopulation from Fig.~2a (main text) and at the coexistence fixed point from the deterministic differential equations (main text eq.~2 or eq.~\ref{eq:diffeq2}). We averaged over all our simulation runs for mixing rates $\mu\,{\gg}\, 50$ in order to obtain a mean for the metapopulation simulations (shading shows the 95\% confidence interval). The deviation from the deterministic differential equations are due to stochastic finite size effects.
\label{f:suppstoch}}
\end{figure}

The Jacobian of the differential equations of eq. 1 in the main text is a $2N \times 2N$ dimensional matrix. However, at these fixed points, the matrix is sparse: as an example, this Jacobian for  $N=2$ (more manageable than $N=6$) evaluated at the fast growing fixed point ($x=0$) reads  
\begin{align*}
	\left( {\begin{array}{cccc}
   f_{1}^{-}-f_{1}^{+} & \omega & 0 & 0 \\
  0 & f_{2}^{-} - f_{1}^{+} -\omega &  0 & 0 \\
		-\omega & -\omega & -f_{1}^{+} & \omega  -f_{2}^{+} \\
   \omega & \omega & 0 & f_{2}^{+}-f_{1}^{-} -\omega \\
	\end{array} } \right) \textrm{.}
\end{align*}
For the fast growing fixed point $x=0$, the derivative of the update terms of the differential equation for fitness types $x_i$ with respect to all fast-growing fitness types $y_i$ are zero because they all involve terms of $x_i$. Similarly, derivatives of the reproduction terms with respect to fitness types $y_i$ retain terms containing $x_i$, and are thus also all zero. The eigenvalues of the Jacobian are thus the eigenvalues of the blocks corresponding to fast growers and slow growers separately (as we know by properties of block determinants that the eigenvalues of a matrix of the form $A=\bigl( \begin{smallmatrix}B & 0\\ C & D\end{smallmatrix}\bigr)$ are the eigenvalues of $B$ and $D$). Analogous arguments show that even these submatrices are  sparse, and that the eigenvalues of these Jacobians are actually the elements on the diagonal. All of these eigenvalues apart from two eigenvalues contain a negative term $\propto \omega$, which for $\omega \gg f$ is negative; of the remaining two eigenvalues, one is also always negative, and the other determines the stability of the fixed point and is discussed and analysed in the main text of the paper. We conclude this section by noting that this last eigenvalue corresponds to the fitness difference between the least fit and fittest individuals of both types (depending on whether the absorbing state of all fast or all slow growers is studied).

We show the concentrations of the individual fitness type  at the coexistence fixed point (for the same parameters as Fig.~2a in main text) as a function of fitness difference $c$ in Fig.~\ref{f:fixedpoint}. The fact that the different fitness types are present at different concentrations shows that the understanding of these different fitness types is important. Finally, we compare the surviving slow-grower fraction at the fixed point of differential equations eq.~2 and eq.~\ref{eq:diffeq2} (obtained by numeric solution using the scipy odeint package) with the Gillespie simulations for the metapopulation from Fig.~2a in Fig.~\ref{f:suppstoch}. The results from the stochastic simulations are closer to the respective absorbing fixed points that the deterministic solution, which is expected due to stochastic finite size effects. Otherwise, the good agreement from the well-mixed deterministic differential equations and the stochastic situation shows that spatial correlations only play a minor role for understanding coexistence in the limit of high mixing rates. 

\section{Phase diagram}
\begin{figure}
        \includegraphics[width=0.8\columnwidth]{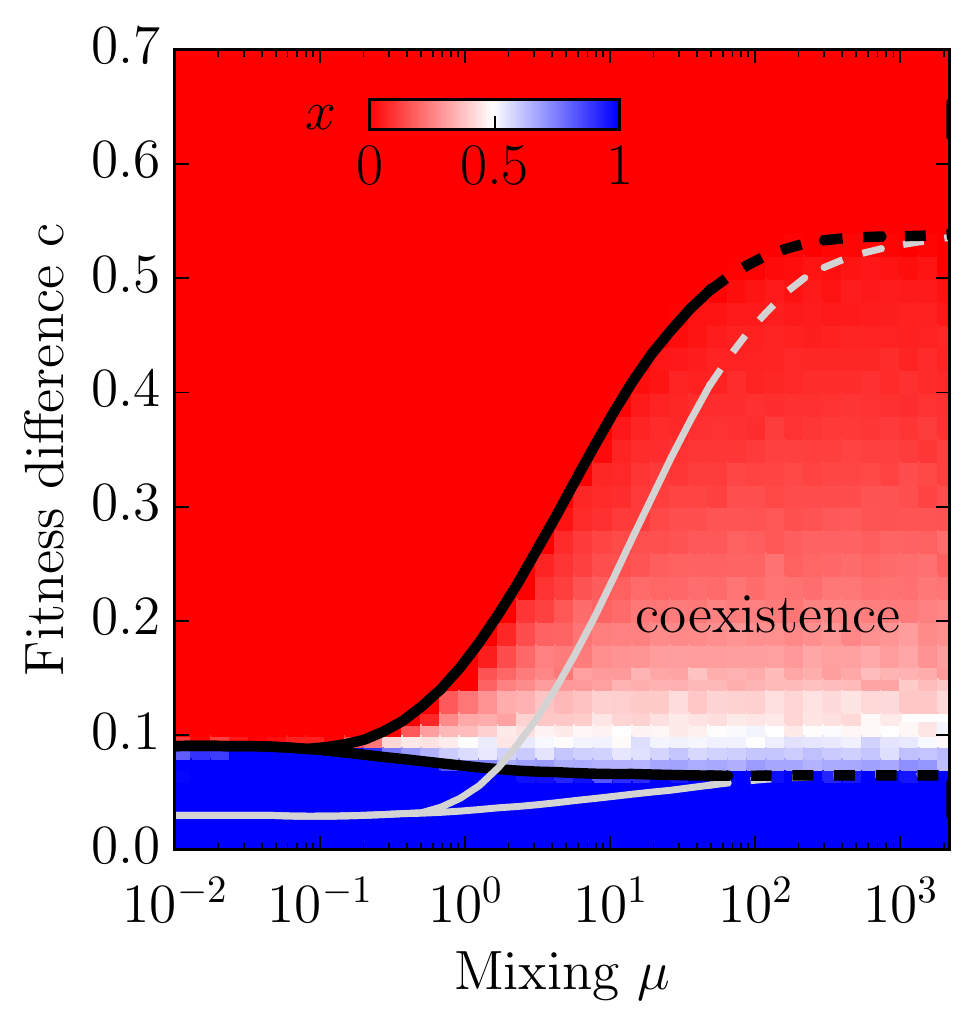}
	\caption{ 	
	Phase diagram from Fig.~2a ($L\times L =30$, $\omega=5$, $N=6$, $\alpha=0.25$), showing the phase boundaries for $\omega=5$ and $\omega=20$ from the extinction time criterion $T_\textrm{ext}=7500$ in solid lines (dashed where stochastic effects are assumed to matter) and colour shading indicating the slow grower density at the last point of the simulations, i.e. either when these converged (outside the coexistence region) or were terminated at $T_\textrm{ext}=25000$ (inside the coexistence region). For high costs, the slow grower fraction becomes small (see Fig.~\ref{f:suppstoch}), yet the stochastic runs still oscillate close to the well-mixed fixed point concentrations for simulation run times $T> T_\textrm{ext}$ (see Fig.~\ref{f:stochruns}).
	\label{f:phasediag}}
\end{figure}

\begin{figure}
        \includegraphics[width=0.95\columnwidth]{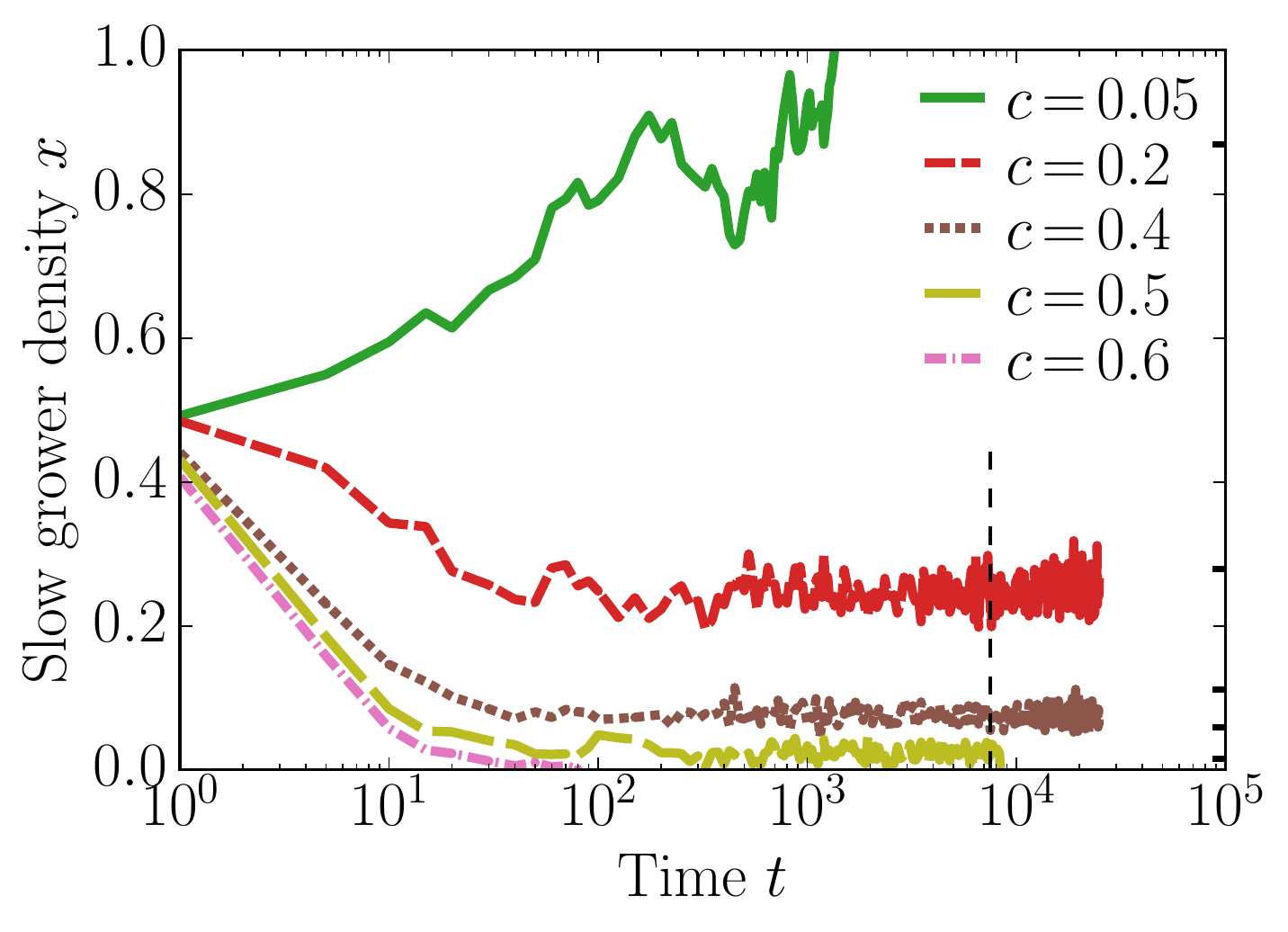}
	\caption{ 	
	Slow grower density $x$ in the metapopulation as a function of simulation time for representative Gillespie simulation runs for $\mu=50$ and five exemplary fitness differences $c$, otherwise same parameters as for Fig.~\ref{f:phasediag}. Our criterion marking coexistence $T_\textrm{ext}=7500$ is marked with the vertical black line. For $c=0.05$ and $c=0.6$ the simulation runs converge fast to the respective stable absorbing state, while the simulation runs for $c=0.2$ and $c=0.4$ show clearly that the slow growers and fast growers coexist. The phase transition between coexistence and the absorbing states of fast growers according to the coexistence criterion is at $c=0.5$. For comparison, well-mixed limits (which would be expected to match with the stochastic simulations for higher $\mu$, modulo finite size effects) are marked in black on the left margin of the images.
	\label{f:stochruns}}
\end{figure}

Fig.~2a in the main text shows the phase diagram of slow and fast grower stability and coexistence based on extinction times in simulation runs. We have verified that these coexistence times are indeed a meaningful quantifier for the different phases. For better intuition on why this is so, we show the phase diagram with the population fraction of slow growers at the end of the simulation run for the exact same parameters ($L\times L =30$, $\omega=5$, $N=6$, $\alpha=0.25$) in  Fig.~\ref{f:phasediag}. The phase boundaries are the same as those from Fig.~2a (obtained from the average extinction times) and show good agreement with the phase boundaries suggested by  Fig.~\ref{f:phasediag}.  We note that as discussed before (cf. in Fig.~\ref{f:suppstoch} and Fig. 2d), the species composition at the coexistence fixed point shifts progressively to fast growers for increasing fitness difference, which is why the population fraction in Fig.~\ref{f:phasediag} appears red even within the coexistence region. 

\begin{figure*}
        \includegraphics[width=0.8\textwidth]{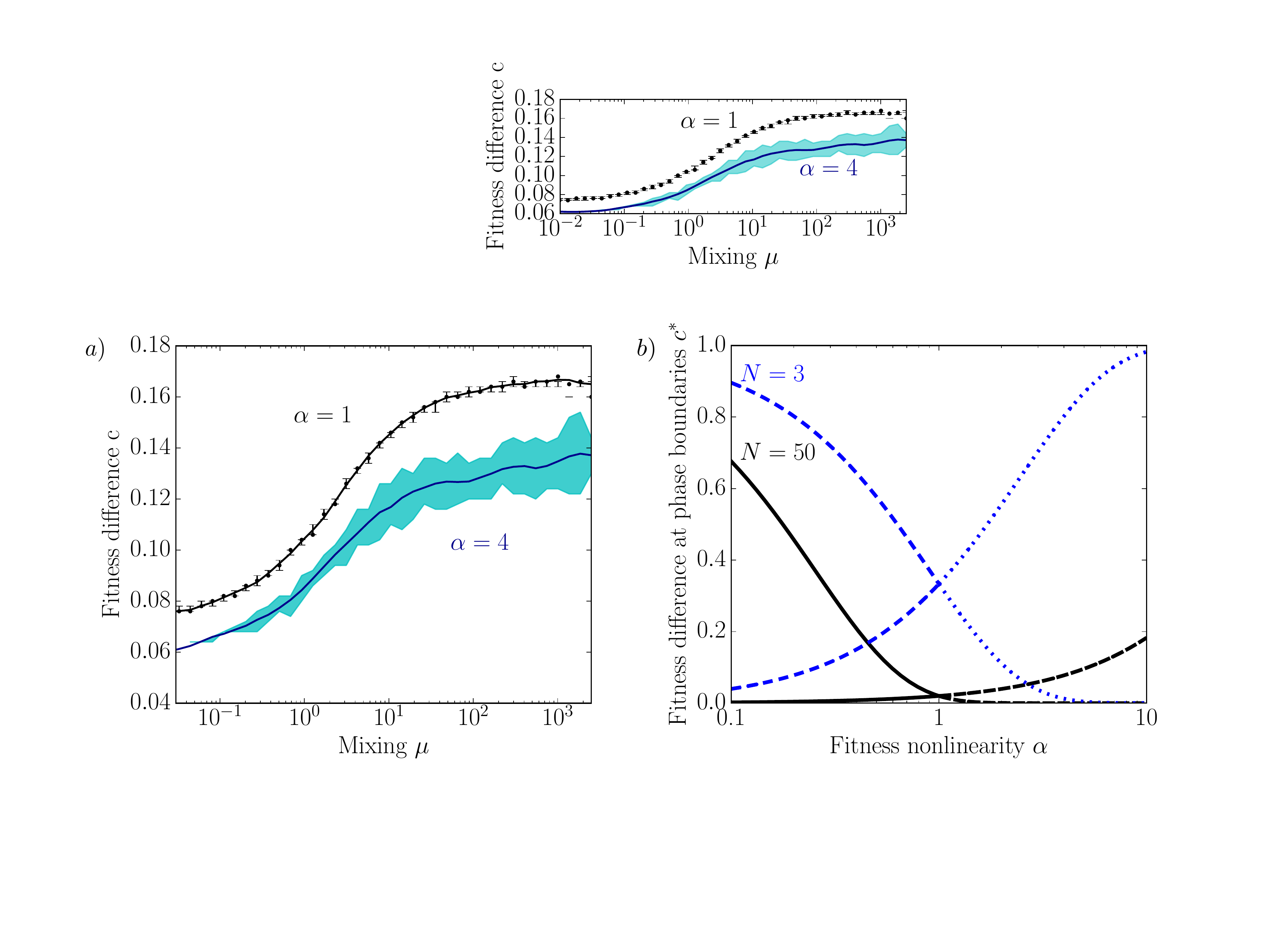}
	\caption{\label{f:bla} 	
	(a) Phase diagrams for $\alpha \,{=}\, 1$ and $\alpha>1$ for comparison with Fig.~2a of the main text. Both lines show a transition between phase where the slow and one where the fast grower is stable (above the line). For $\alpha \,{=}\, 1$, the data for both phase boundary and error bars is taken from Ref.~[78] in the main text. The bistability for  $\alpha>1$, here shown for a representative value of $\alpha \,{=}\, 4$ means that the outcome depends on initial conditions. This dependence also means that there are large fluctuations around the phase boundary, which we have indicated by the blue shading. 
	(b) The value of fitness difference at the phase boundary, $c^{*}$, between stable slow grower and coexistence ($c^{-}$  in the main text) and stable fast growers and coexistence ($c^{+}$), for $N \,{=}\, 3$ and $N \,{=}\, 50$, for comparison with Fig.~2d of the main text for $N \,{=}\, 6$.
	}
\end{figure*}

In order to illustrate why we can characterise different phases in terms of average extinction time, we show exemplary simulation runs for $\mu=50$ and five exemplary fitness differences in Fig.~\ref{f:stochruns}: it is clear from this figure that for fitness differences in the coexistence region ($c=0.2$ and $c=0.4$), the slow grower density oscillates about the respective coexistence fixed point.  Our criterion for the coexistence phase, $T_\textrm{ext}=7500$, suggests that the  phase transition between coexistence and the absorbing states of fast growers  is at $c=0.5$ for $\mu=50$. We expect that the average extinction time for systems with parameters close to the boundary of the coexistence phase  will increase with system size, as both the impact of spatial correlations as well as finite size effects will diminish for larger system sizes. Indeed, for $\mu\ge 50$, the phase boundary shown in Fig.~\ref{f:phasediag} is marked in dashes, as it is susceptible to changes in the (large) extinction time criteria or the system size, due to these correlations and finite size effects. Conversely, for $\mu\le 50$, the phase boundary does not vary significantly with changes of extinction time criteria or system size, and is thus marked with solid lines in Fig.~\ref{f:phasediag}. A more detailed study of the phase boundaries would go beyond the scope of this work, as we are predominately interested in a proof of principle of existence of the coexistence phase, and its robustness over a significant range of different mixing rates and fitness differences.

\section{Impact of nonlinearity coefficient and $N$}

The phase boundary for $\alpha \,{=}\, 1$ in Fig.~\ref{f:bla}a (in black) shows that the region where the slow-grower is stable goes up to $c^{*} \,{=}\, 1/N$ in the well-mixed limit. For $\alpha \,{>}\, 1$, the phase boundary in the well-mixed limit is not well-defined (i.e. there is no phase transition in the system), as the system is bistable and thus depends strongly on initial conditions. The region where between $30\%$ and $70\%$ of all simulation runs converge to slow growers is shown in light blue. Within this blue region, each simulation run converges to either slow or fast growers, but not to no coexistence. The average position of the phase boundary is shown in dark blue as a guide to the eyes. We note that the phase boundary in the small mixing limit changes quantitatively but not qualitatively with both $\alpha$ and $\omega$.

Figure~\ref{f:bla}b shows the boundaries for the fixed point stability $c^{+} \,{=}\, 1/N^\alpha$ and $c^{-} \,{=}\, 1-((N-1)/N)^\alpha$ for $N \,{=}\, 50$ (black, solid and long dashes) and $N \,{=}\, 3$ (blue, short dashes and dashed dotted). Even for increased $N$ the coexistence region (between the lines) extends over a large range of fitness difference $c$. This fact means that coexistence is also robust even for higher $N$, especially when one considers that the fitness function may be even more nonlinear (decreasing $\alpha$). In addition, when it comes to debating the extent of different phases, our explanation of how the different fitness types needs to balance each other makes clear that the importance of $N$ for the phase stability is that it sets the difference between the least and most fit individuals of both species. This fitness difference could (for larger populations) also be an effect observed due to private components of the public good, as discussed in the main text. Because different experimental cases would require different fitness functions and more specific modeling, it is reasonable to limit the discussion of the point of the importance of $N$ in our model to only providing evidence for the fact that coexistence also also arises and is robust for larger $N$.

\end{document}